\documentclass[conference]{IEEEtran}
\IEEEoverridecommandlockouts

\usepackage{cite}
\usepackage{amsmath,amssymb,amsfonts}
\usepackage{algorithmic}
\usepackage{graphicx}
\usepackage{textcomp}
\usepackage{xcolor}
\def\BibTeX{{\rm B\kern-.05em{\sc i\kern-.025em b}\kern-.08em
		T\kern-.1667em\lower.7ex\hbox{E}\kern-.125emX}}

\usepackage{multirow}
\usepackage{upgreek}
\usepackage{tabularx}
\usepackage{threeparttable}
\usepackage[font=small]{caption}
\usepackage{subcaption}
\newcolumntype{Y}{>{\centering\arraybackslash}X}
\usepackage{amsthm}

\usepackage{enumitem}
\usepackage[ruled, linesnumbered]{algorithm2e}

\newcolumntype{C}{>{\centering\arraybackslash}X}

\DeclareMathSymbol{\shortminus}{\mathbin}{AMSa}{"39}

\newcommand{\trace}{\mathrm{tr}}
\newcommand{\rank}{\mathrm{rank}}

\newcommand{\exponent}{\mathrm{exp}}
\newcommand{\vectorize}{\mathrm{vec}}

\newcommand{\conn}{\mathtt{C}}

\newcommand{\fd}{\mathit{\Delta f}}
\newcommand{\fc}{f_{\mathrm{c}}}

\newcommand{\SINRcomm}{\mathrm{SINR}^{\mathrm{com}}}
\newcommand{\SINRcomms}{\overline{\mathrm{SINR}}^{\mathrm{com}}}

\newcommand{\Tsym}{T_{\mathrm{sym}}}

\newcommand{\Tdft}{T} %_{\mathrm{dft}}

\newcommand{\Tgi}{\mathit{T_{\mathrm{cp}}}}

\newcommand{\rangeRes}{\mathit{\Delta R}}

\newcommand{\angleRes}{\mathit{\Delta \theta}}

\newcommand{\rangeMax}{R_{\mathrm{max}}}

\newcommand{\snrradar}{\mathrm{SNR}^{\mathrm{rad}}}
\newcommand{\targetrcs}{\mathit{\sigma_{\mathrm{rcs}}}}

\newcommand{\mydots}{\!...}
\newcommand{\Tstrut}{\rule{0pt}{2.6ex}}

\newcommand{\Mtarget}{M_{\mathrm{trgt}}}

\newcommand{\Nang}{N_{\mathrm{ang}}}

\newcommand{\Nsym}{N_{\mathrm{sym}}}

\newcommand{\Ntx}{N_{\mathrm{tx}}}
\newcommand{\Nvirt}{N_{\mathrm{virt}}}
\newcommand{\Dvirt}{D_{\mathrm{virt}}}

\newcommand{\Nrx}{N_{\mathrm{rx}}}
\newcommand{\Mrx}{M_{\mathrm{rx}}}
\newcommand{\Nsc}{N_{\mathrm{sc}}}

\newcommand{\Dsc}{D_{\mathrm{sc}}}
\newcommand{\Ptx}{P_{\mathrm{tx}}}
\newcommand{\Grx}{G_{\mathrm{rx}}}

\newcommand{\radarChan}{\textbf{H}^{\mathrm{rad}}}
\newcommand{\radarChanEst}{\hat{\textbf{H}}^{\mathrm{rad}}}
\newcommand{\radarChanVec}{\textbf{h}^{\mathrm{rad}}}

\begin{document}

\title{Optimal Precoder Design for MIMO-OFDM-based Joint Automotive Radar-Communication Networks \thanks{This work has been supported in part by NSF through grants 1814923, 1955535.}
}

\author{\IEEEauthorblockN{Ceyhun D. Ozkaptan, Eylem Ekici}
	\IEEEauthorblockA{Dept. of Electrical and Computer Engineering \\
		The Ohio State University\\
		Columbus, OH, USA \\
		\{ozkaptan.1, ekici.2\}@osu.edu}
	\and
	\IEEEauthorblockN{Chang-Heng Wang, Onur Altintas}
	\IEEEauthorblockA{InfoTech Labs\\
		Toyota Motor North America \\
		Mountain View, CA, USA \\
		\{chang-heng.wang, onur.altintas\}@toyota.com}
	}

\maketitle

\begin{abstract}
	
Large-scale deployment of connected vehicles with cooperative awareness technologies increases the demand for vehicle-to-everything (V2X) communication spectrum in 5.9 GHz that is mainly allocated for the exchange of safety messages. To supplement V2X communication and support the high data rates needed by broadband applications, the millimeter-wave (mmWave) automotive radar spectrum at 76-81 GHz can be utilized. For this purpose, joint radar-communication systems have been proposed in the literature to perform both functions using the same waveform and hardware. While multiple-input and multiple-output (MIMO) communication with multiple users enables independent data streaming for high throughput, MIMO radar processing provides high-resolution imaging that is crucial for safety-critical systems. However, employing conventional precoding methods designed for communication generates directional beams that impair MIMO radar imaging and target tracking capabilities during data streaming. In this paper, we propose a MIMO joint automotive radar-communication (JARC) framework based on orthogonal frequency division multiplexing (OFDM) waveform. First, we show that the MIMO-OFDM preamble can be exploited for both MIMO radar processing and estimation of the communication channel. Then, we propose an optimal precoder design method that enables high accuracy target tracking while transmitting independent data streams to multiple receivers. The proposed methods provide high-resolution radar imaging and high throughput capabilities for MIMO JARC networks. Finally, we evaluate the efficacy of the proposed methods through numerical simulations.

\end{abstract}

\section{Introduction}

As an integral part of Intelligent Transportation Systems (ITS), the connected vehicle technology will promote safer and coordinated transportation through wireless communication and sensing technologies. To enable vehicle-to-everything (V2X) communication, Federal Communications Commission (FCC) dedicated 30 MHz in the 5.9 GHz band for the ITS applications and the exchange of safety-related messages. With the large-scale deployment of connected vehicles, the V2X spectrum in the 5.9 GHz band will face a spectrum scarcity problem and will not be able to sustain non-safety-related and broadband applications due to limited bandwidth. Moreover, emerging cooperative sensing and full self-driving technologies may require a large amount of raw sensor and navigation data to be exchanged for improved reliability and performance \cite{2015_cooperative_perception}. However, the V2X spectrum cannot be used efficiently for larger payloads along with basic safety messages that are crucial for safety-critical applications. A solution to alleviate the scarcity problem and attain higher data rates is to leverage the underutilized millimeter-wave spectrum (mmWave).

Meanwhile, 76-77 GHz and 77-81 GHz mmWave spectra are dedicated to the automotive long-range radar (LRR) and short-range radar (SRR) operations with the contiguous bandwidths of 1 and 4 GHz, respectively. Since automotive radars are crucial for assisted driving and safety-critical systems (e.g., adaptive cruise control, collision avoidance), large bandwidths with high transmit power limits are provided in the mmWave spectrum to enable better estimation accuracy of targets' distance, velocity, and angle \cite{2012_77ghz_radar_sensors}. Therefore, one prominent solution to supplement the V2X communication and support a high data rate is the deployment of joint radar-communication systems in 76-81 GHz automotive radar band for simultaneous radar imaging and data transmission. 

Over the past decade, numerous waveform design and processing methods have been proposed for joint radar-communication systems. Linear frequency modulated (LFM) waveform, which is the most commonly used waveform in radar systems, is investigated as a dual-use waveform. While some of the proposed approaches have leveraged spread-spectrum methods \cite{css_beijing_2011} for simultaneous transmission, other studies have employed phase-coding methods \cite{nowak_lfm_2016} to encode data on LFM waveform. However, proposed solutions offer low data rates and compromise both communication and radar performance due to interference and arbitrary data encoding. Radar processing methods that employ \textit{conventional communication waveforms} have also been proposed in the literature to avoid degradation of the communication performance. In \cite{heath_802.11ad_radar_2018}, IEEE 802.11ad-based radar processing has been investigated to utilize the preamble of single-carrier indoor communication standard for range and velocity estimation. Since a single-carrier waveform designed for indoor communication is employed, the proposed system lacks the resilience against mobility and frequency-selective fading. Besides single-carrier communication waveform, several OFDM-based radar processing methods are proposed as a joint system in \cite{garmatyuk_ofdm_2009, sturm_ofdm_2011, ozkaptan_ofdm_pilot, infocom_2020}. Correlation-based radar processing has been investigated with the OFDM waveform in \cite{garmatyuk_ofdm_2009, infocom_2020}. A symbol-based processing method has been proposed for in \cite{sturm_ofdm_2011} to estimate range and velocity based on the phase shift on modulation symbols. Moreover, in \cite{ozkaptan_ofdm_pilot}, the pilot sequences in OFDM waveform are used for radar processing and channel estimation. Among these approaches, OFDM-based systems are attractive candidates for joint automotive radar-communication systems considering broadband capabilities such as robustness against frequency-selectivity and low complexity equalization.

Today, automotive radar systems are equipped with multiple transmit and receive antennas to enable beamforming and multiple-input and multiple-output (MIMO) radar processing capabilities. In recent years, various multi-antenna joint systems are investigated for iterative parameter estimation \cite{fan_liu_joint_design_2020}, and transmit beampattern design \cite{fan_liu_joint_design_2018} without actual radar imaging methods. In this work, we propose a MIMO-OFDM-based joint radar-communication system that enables multi-user communication and high accuracy radar imaging. Different from previous work in the literature, the proposed precoder design method provides simple yet effective beamforming and radar imaging framework that can achieve high accuracy target tracking while transmitting data streams to multiple receivers. Moreover, the proposed framework can be adapted to work with the IEEE 802.11n/ac standards that employ the MIMO-OFDM waveform. The main contributions of this work are summarized as follows:
\begin{itemize}[leftmargin=*]
\item  We propose an efficient range-angle imaging method based on the Fourier transform that leverages both the orthogonal preamble and precoded symbols of MIMO-OFDM waveform for MIMO radar imaging with high-resolution. 
\item We derive the performance metrics for the proposed MIMO radar imaging and multi-user MIMO communication functionalities based on positioning accuracy and communication signal-to-interference-plus-noise ratio (SINR).
\item We propose a joint precoder design method that enables independent data transmissions with \textit{high data rates} while steering the remaining transmit power for tracking multiple targets with \textit{high positioning accuracy}. Moreover, we evaluate the trade-offs between communication and radar performance in a vehicular scenario through simulations.

\end{itemize}

\textbf{Notations}: Boldface lower-case letters denote column vectors $ \textbf{a} $, boldface upper-case letters denote matrices $ \textbf{A} $, and plain lower-case letters denote scalars $ a $. $ \mathbb{R} $, and $ \mathbb{C} $ define the sets of real numbers, and complex numbers, respectively. The superscripts $ |\cdot| $, $ (\cdot)^* $, $ (\cdot)^T $, and $ (\cdot)^{\dagger} $ denote the absolute value, the complex conjugate, transpose, and the transpose conjugate, respectively. $\textbf{A}[m,n] $ indicates the element in row $ m $ and column $ n $ of matrix $ \textbf{A} $. For slicing the matrices, $ [m:n] $ and $ [:] $ denote the indices from $ m $ to $ n $ and all indices along the dimension, respectively. 

\section{System Model}\label{section_system_model}

\begin{figure}[!t]
	\centering
	\includegraphics[width=0.8\linewidth]{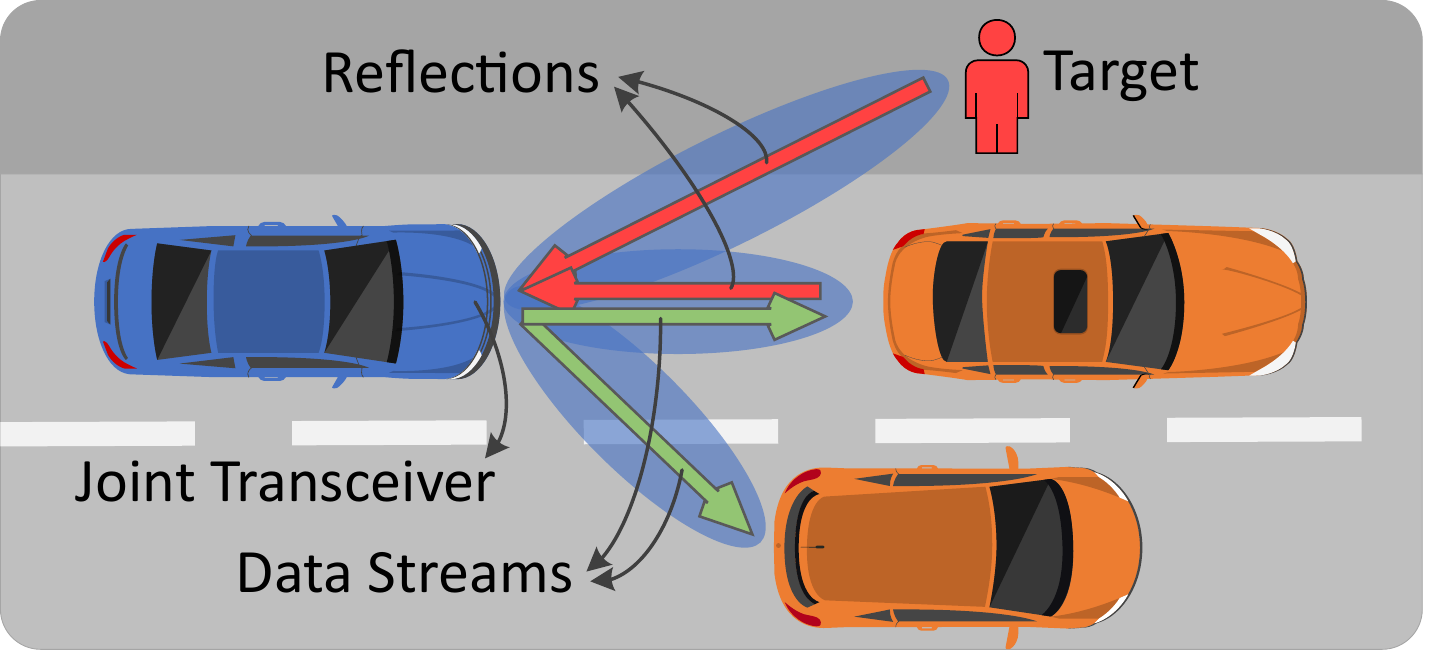}
	\caption{A vehicular scenario, where a joint radar-communication capable vehicle (Blue) generates beams for simultaneous target tracking and data streaming.}
	\label{fig_system}
\end{figure}

In this section, we describe our system model for the joint MIMO system by formulating the transmitted MIMO-OFDM signal and outlining the transceiver architecture for virtual MIMO radar processing. As illustrated in Fig.~\ref{fig_system}, the transmitter vehicle sends independent data streams with MIMO-OFDM waveform to receiver vehicles. At the same time, the transmitter vehicle also operates as the radar receiver (i.e., joint transceiver) that processes the reflected signals to generate a range-angle image (i.e., radar image) of the illuminated region. 

\begin{figure}[!b]
	\centering
	\includegraphics[width=0.9\linewidth]{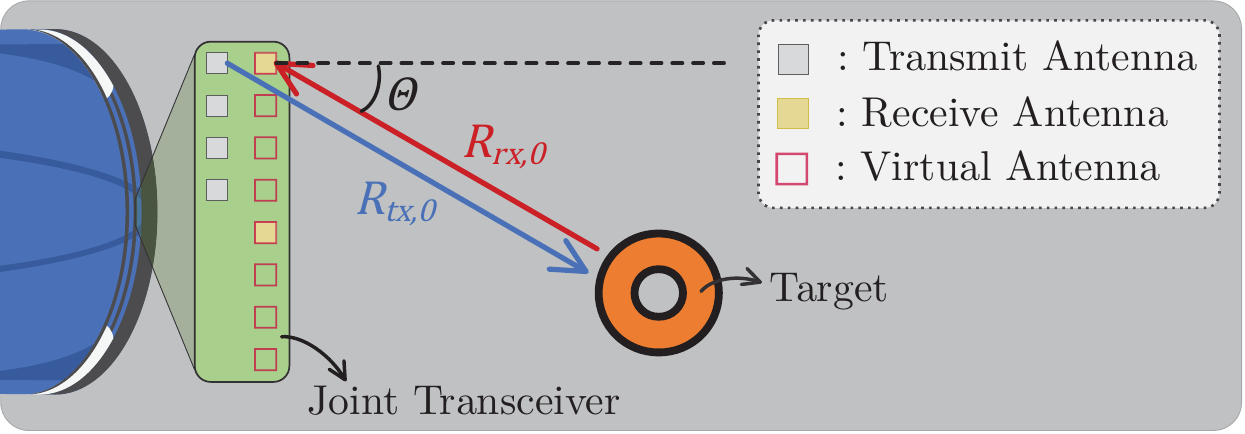}
	\caption{A joint MIMO transceiver with a virtual ULA with  $ \Nvirt=8 $ antennas formed with $ \Ntx=4 $ transmit and $ \Nrx=2 $ receive antennas.}
	\label{fig_antenna}
\end{figure}

We consider a joint MIMO-OFDM system that uses a uniform linear array (ULA) comprising $ \Ntx $ transmit antennas with critical spacing of $ d_{\text{tx}}=\lambda/2 $ to achieve full angular resolution, where $ \lambda = c/\fc  $ is the wavelength, $ c $ is the speed of light, and $ \fc $ is the carrier frequency. As depicted in Fig.~\ref{fig_antenna}, the system is also equipped with a receive ULA comprising of $ N_{\text{rx}} $ antennas with sparse spacing of $ d_{\text{rx}}=\Ntx\lambda/2 $ to form a virtual aperture with MIMO radar processing. The combined response of critically-spaced $ \Ntx $ transmit and sparsely-spaced $ N_{\text{rx}} $ receive antennas is equivalent of a virtual ULA with $ \Nvirt = \Ntx\Nrx $ elements that achieves higher degrees of freedom and angular resolution of $ \angleRes = 2/\Nvirt $ radians \cite{li2009mimo}. In this work, we consider radar processing with ULAs for 2D imaging in range and azimuth angle domain for simplicity and brevity. Nevertheless, the principles proposed in this work can be extended to 3-dimensional (3D) radar imaging and beamforming by using 2-dimensional (2D) antenna arrays such as uniform rectangular arrays (URA). 

\subsection{Transmitted Signal}

For optimal MIMO radar processing, the transmissions from $ \Ntx $ antennas should be orthogonal (i.e. uncorrelated) to separate the signals from different transmit antennas \cite{li2009mimo}. Similarly, optimal estimation of the MIMO communication channel also requires orthogonal transmission that can be achieved with the fixed preamble symbols. Hence, an orthogonal preamble is added for every transmission for both MIMO radar processing and communication channel estimation. At the $ l^{\text{th}} $ transmit antenna, precoded MIMO-OFDM signal with $ \Nsc $ subcarriers, $ \Nsym $ symbols, and $ \Ntx $ spatial streams is generated in the continuous-time domain as 
\begin{equation}
	\label{eqn_tranmit_signal}
	\begin{aligned}
		\tilde{x}_{l} (t) = \sum_{m=0}^{\Nsym\!\shortminus 1}\sum_{n=-\Nsc/2}^{\Nsc/2\!\shortminus1} & \textbf{F}_n[l,:] \textbf{S}_n[:,m]   \times \exponent\big(j2\pi n \fd t\big) \\
		& \times g(t-m \Tsym),
	\end{aligned}
\end{equation}
where $ \textbf{F}_n \in \mathbb{C}^{\Ntx \times \Ntx}  $ is the joint precoding matrix, and $ \textbf{S}_n \in \mathbb{C}^{\Ntx \times \Nsym}  $ contains independent data and radar symbols with unit energy. Based on \eqref{eqn_tranmit_signal}, the orthogonal preamble symbols are generated with $ \Nsym = \Ntx$, $ \textbf{F}_n = \sqrt{P_{\mathrm{tx}}/\Ntx} \textbf{I}_{\Ntx} $, and $ \textbf{S}_n = \textbf{P}_n$, where $ P_{\mathrm{tx}} $ is the total transmit power, $ \textbf{P}_n \in \mathbb{C}^{\Ntx \times \Ntx}  $ is the orthogonal mapping matrix that is $ \textbf{P}_n\textbf{P}_n^{\dagger} = \Ntx \textbf{I}_{\Ntx} $.

For the orthogonality of subcarriers, the frequency spacing is set to $\fd = 1/\Tdft$, where $ \Tdft $ is the OFDM symbol duration without cyclic prefix and the total bandwidth of signal is $ B = \Nsc\fd $. To avoid inter-symbol interference (ISI) and maintain circular convolution with the channel, the cyclic prefix is prepended to each OFDM symbol with the rectangular pulse shaping function $ g(t) = 1/\sqrt{\Nsc}$ for $t\in[-\Tgi, \Tdft]$, where $ \Tsym = \Tdft + \Tgi $ is the total OFDM symbol duration with cyclic prefix. For the transmission, the baseband signal is upconverted to the carrier frequency $ \fc $ and transmitted via ULA with $ \Ntx $ antennas, which is expressed in vector form as 
	$\tilde{\textbf{x}}_{\text{rf}}(t) = \mathrm{Re}\left\{\tilde{\textbf{x}}(t)\exp(j2\pi \fc t)\right\},$
where $  \tilde{\textbf{x}}(t) = [\tilde{x}_{0}(t),\mydots, \tilde{x}_{\Ntx\!\shortminus\!1} (t) ]^T $.

\subsection{Communication Channel}

For the communication channel, we consider a multi-user scenario with $ \Mrx $ receivers that employ single antennas for data reception where $ \Mrx < \Ntx $. When no communication link is established, the joint transceiver \textit{periodically} transmits \textit{null data packets} (NDP) that only contain the orthogonal preamble for MIMO radar imaging that generates an omnidirectional transmission pattern. Besides MIMO radar imaging functionality, the NDPs are also used as channel sounding packets for communication where NDP receivers estimate the communication channels with the known preamble. 

At $ q^{\text{th}} $ communication receiver, the received preamble symbols in $ n^\text{th} $ subcarrier are denoted by vector $ \textbf{y}^{\mathrm{(com)}}_{n,q} \in \mathbb{C}^{\Ntx}$ and formulated as 
\begin{equation}
	\label{eqn_rx_comm_signal}
	{\textbf{y}^{\mathrm{com}}_{n,q}}^T = {\textbf{h}^{\mathrm{com}}_{n,q}}^T \textbf{P}_n + {\textbf{w}^{\mathrm{com}}_{n,q}}^T, 
\end{equation}
for $ n = -\Nsc/2,\mydots,\Nsc/2\!\shortminus\!1 $ and $ p=0,\mydots,\Ntx\!\shortminus\!1$, where $ q \in \Psi_{\mathrm{com}} $ is the set of indices of $\Mrx$ receivers, $ {\textbf{h}^{\mathrm{com}}_{n,q}} \in \mathbb{C}^{\Ntx } $ is the communication channel, and $ {\textbf{w}^{\mathrm{com}}_{n,q}} $ is zero-mean complex Gaussian noise with covariance $ \sigma^2_{\mathrm{com},q}\textbf{I}_{\Ntx} $. With the known preamble symbols in $ \textbf{P}_n $, the receiver estimates the channel $ {\textbf{h}^{\mathrm{com}}_{n,q}} $ as shown in \cite{perahia2013next}. Then, the receiver feedbacks the estimates of channel $ \hat{\textbf{h}}^{\mathrm{com}}_{n,q} $ and noise variance $ \hat{\sigma}^2_{\mathrm{com},q} $  to the transmitter as channel state information (CSI) for precoder design and spatial multiplexing. 

Based on the precoder and the structure of the multi-user MIMO channel, data streams for other users and radar streams can cause interference on $ q^{\text{th}} $ receiver's data stream. Since SINR determines the communication channel capacity of each receiver \cite{tse_viswanath_2005}, the joint transmitter ensures that the received SINR of each data stream are above a certain limit while designing the joint precoder. Thus, the SINR constraint for the data stream of $q^{\text{th}} $ receiver in $ n^\text{th} $ subcarrier is defined as
\begin{equation}
	\label{eqn_comm_sinr}
	\SINRcomm_{n,q}\left(\textbf{f}_{q,n}\right) = \frac{ \left|\hat{\textbf{h}}^{\mathrm{com}\dagger}_{n,q} \textbf{f}_{q,n}\right|^2}{\sum_{j=1, j\neq q}^{\Mrx}\left|\hat{\textbf{h}}^{\mathrm{com}\dagger}_{n,q} \textbf{f}_{j,n}\right|^2 +  \hat{\sigma}^2_{\mathrm{com},q}} \geq \eta_q,
\end{equation}
where $ \textbf{f}_{q,n} $ is the $q^{\text{th}} $ column of joint precoder $ \textbf{F}_n $, and $ \eta_q $ is the minimum SINR requirement to support chosen modulation scheme with a low error rate.

\subsection{Radar Channel}

We consider radar channel contains $ \Mtarget $ point targets with the gains $\{\alpha_p\}_{p \in \Psi_{\mathrm{rad}}} \in \mathbb{C}$ (i.e., contains path loss and reflectivity) that are located $ \{R_p\}_{p \in \Psi_{\mathrm{rad}}}$ meters away from the joint transceiver at an azimuth angle of $ \{\varTheta_p\}_{p \in \Psi_{\mathrm{rad}}} $, where $ \Psi_{\mathrm{rad}} $ is the set of indices of $\Mtarget$ targets. For brevity, we first model the reflected signal from the $p^{\text{th}} $ target as the radar channel will be the superposition of $ \Mtarget $ reflected signals. After the transmitted signal is reflected from the target, it is received by the receive ULA with $ \Nrx $ antennas and downconverted to the baseband. Since the total duration of radar illumination is short (i.e., $ \Nsym\Tsym $), we assume the target is quasi-stationary and its parameters are constant. After the downconversion, the baseband signal that is received by the $ k^{\mathrm{th}} $ receive antenna is expressed in the continuous-time domain as  
\begin{equation}
	\label{eqn_rx_signal}
	\begin{gathered}
		\tilde{y}^{\mathrm{rad}}_k(t)  = \sum_{l=0}^{\Ntx\!\shortminus1} \alpha_p \tilde{x}_{l}(t-\tau_{k,l}) \exp\big(\!-\!j 2 \pi \fc \tau_{k,l}\big) + \tilde{w}^{\mathrm{rad}}_k(t),\\
	\end{gathered}
\end{equation}
where $ \tilde{w}^{\mathrm{rad}}_k(t) $ is the additive complex Gaussian noise. The target's gain $ \alpha_p $ is defined in \cite{book_richards_radar_principles_2010} as
\begin{equation}
	\alpha_p = \sqrt{\frac{\targetrcs_{,p} \Grx \lambda^2 }{(4\pi)^3 R_p^4}}\exp(j\phi_p),
	\label{eqn_attenuation}
\end{equation} 
where $ \Grx $ is the receive gain, $\targetrcs_{,p}$ is the reflectivity and $ \phi_p $ is the phase of the $p^{\text{th}} $ target. Also, $ \tau_{k,l,p} = D_{k,l,p}/c$ in \eqref{eqn_rx_signal} denotes the two-way propagation delay, where $ c $ is the speed of light, $ D_{k,l,p}$ is the total distance traveled by the signal that is the sum of the distances from the target to $ l^{\mathrm{th}} $ transmit antenna and to $ k^{\mathrm{th}} $ receive antenna denoted by $  R_{\mathrm{tx},l,p} $ and $ R_{\mathrm{rx},k,p} $, respectively.

After sampling the continuous-time received signal in \eqref{eqn_rx_signal} at $ t = i\Tdft/\Nsc + m\Tsym $, the discrete-time baseband signal is represented in matrix form denoted by $ \tilde{\textbf{Y}}^{\mathrm{rad}}_k \in \mathbb{C}^{\Nsc \times \Nsym} $ whose elements are formulated as 
\begin{equation}
	\label{eqn_rx_signal_sampled}
	\begin{gathered}
		\begin{aligned}
			\tilde{\textbf{Y}}^{\mathrm{rad}}_k[i,m] & = \alpha_p \sum_{l=0}^{\Ntx\!\shortminus 1}   \sum_{n=-\Nsc/2}^{\Nsc/2\!\shortminus1}  \textbf{F}_n[l,:] \textbf{S}_n[:,m]  \\
			& \times \exponent\big(j2\pi n \fd i \Tdft / \Nsc\big) \exponent\big(\!-\!j2\pi f_n \tau_{k,l,p} \big) \\
		\end{aligned}
	\end{gathered}
\end{equation}
where $ f_n = \fc + n \fd $. In \eqref{eqn_rx_signal_sampled}, the discrete-time baseband signal consists of two exponent terms that correspond to the OFDM modulation (i.e., IDFT) and a phase shift component due to the two-way propagation delay. To remove the OFDM modulation, the discrete Fourier transform (DFT) is first performed on $ \tilde{\textbf{Y}}^{\mathrm{rad}}_k[i,m] $ along $ i $-axis. With the DFT operation, the received signal is formulated in discrete-frequency (i.e., subcarrier) domain as
\begin{equation}
	\label{eqn_rx_signal_freq}
	\begin{gathered}
		\begin{aligned}
			\textbf{Y}^{\mathrm{rad}}_k[n,m]  =  \sum_{i=0}^{\Nsc\!\shortminus1} & \tilde{\textbf{Y}}^{\mathrm{rad}}_k[i,m] \times \exponent\big(\!-\!j2\pi n i / \Nsc\big)\\
			= \sum_{l=0}^{\Ntx\!\shortminus 1} & \alpha_p \textbf{F}_n[l,:] \textbf{S}_n[:,m] \exponent\big(\!-\!j2\pi f_n\tau_{k,l,p} \big),
		\end{aligned}
	\end{gathered}
\end{equation}
for $n = -\Nsc/2,\mydots,\Nsc/2\!\shortminus\!1$, and $ m = 0,\mydots,\Nsym\!\shortminus\!1.$ The reflected symbols in \eqref{eqn_rx_signal_freq} highlight the phase shift effects due to the propagation delay that depends on both carrier frequency $ f_n $ at $ n^\text{th} $ subcarrier and the placement of transmit-receive antenna pairs denoted by indices $ l $ and $ k $, respectively. 

To simplify the notations and formulate received symbols per subcarrier in matrix form, we rearrange matrix dimensions in \eqref{eqn_rx_signal_freq} as
\begin{equation}
	\label{eqn_rx_signal_freq_simp}
	\boldsymbol{\mathcal{Y}}^{\mathrm{rad}}_{n} = \radarChan_{n,p} \textbf{F}_n  \textbf{S}_n + \textbf{W}^{\mathrm{rad}}_{n}, 
\end{equation}
where $ \boldsymbol{\mathcal{Y}}^{\mathrm{rad}}_{n} \in \mathbb{C}^{\Nrx \times \Nsym} $ contains received complex symbols in $ n^\text{th} $ subcarrier with $ \Nrx $ receive antennas and $ \Nsym $ OFDM symbols. Also, $ \textbf{W}^{\mathrm{rad}}_{n} $ is the additive complex noise where $ \vectorize(\textbf{W}^{\mathrm{rad}}_{n}) \sim \mathcal{CN}(\textbf{0},  \sigma^2_{\mathrm{rad}}\textbf{I}_{\Nrx\Nsym})$. Hence, the elements of the MIMO radar channel matrix $ \radarChan_{n,p} \in \mathbb{C}^{\Nrx \times \Ntx} $ are defined as
\begin{equation}
	\label{eqn_radar_chan}
	\radarChan_{n,p}[k,l] = \alpha_p \exponent\big(\!-\!j2\pi f_n\tau_{k,l,p} \big).
\end{equation}

Since transmit and receive antennas are colocated, we assume that the distances from the target to first antenna elements are equal as $ R_{\mathrm{tx},0} = R_{\mathrm{rx},0} = R_p$. In this case, we define the range-dependent steering vectors of the transmit ULA and the receive ULA for the  $ n^\text{th} $ subcarrier as
\begin{equation}
			\setlength{\jot}{5pt}
	\label{eqn_steer_vec1}
	\begin{gathered}
	\begin{aligned}
		\textbf{a}_{\mathrm{tx},n}(R_p,\varTheta_p)[l] &=  \exponent\left(\!-\!j2\pi f_n \big(R_p-l d_{\mathrm{tx}} \sin(\varTheta_p) \big)/c \right), \\
		\textbf{a}_{\mathrm{rx},n}(R_p,\varTheta_p)[k] &=  \exponent\left(\!-\!j2\pi f_n \big(R_p-k  d_{\mathrm{rx}} \sin(\varTheta_p)\big)/c \right), \\
	\end{aligned}
	\end{gathered}
\end{equation}
respectively, based on the target's azimuth angle $ \varTheta_p $ considering that the distance to target is much larger than the antenna array size (i.e., plane wave approximation in far-field) \cite{tse_viswanath_2005}.

Given the definitions of steering vectors and the channel response for target $ p $, the MIMO radar channel is the superposition of responses from $ \Mtarget $ targets which is expressed as 
\begin{equation}
	\label{eqn_radar_chan3}
	\radarChan_{n} = \sum_{p=1}^{\Mtarget}\! \radarChan_{n,p} = \sum_{p=1}^{\Mtarget}\! \alpha_p \textbf{a}_{\mathrm{rx},n}(R_p,\varTheta_p) \textbf{a}_{\mathrm{tx},n}(R_p,\varTheta_p) ^T 
\end{equation} 
for $ \Mtarget $ targets. Nonetheless, the detection and separation capabilities of multiple targets depend on the range and angle resolution of the radar which are explained in the next section.

\section{Range-Angle Imaging Method}\label{section_range_angle}

In this section, we present the radar processing methods to generate the range-angle images of the illuminated region by leveraging the reflected symbols in the MIMO-OFDM waveform defined in \eqref{eqn_rx_signal_freq_simp}. The received signal samples are the superposition of precoded symbols transmitted from different antennas on different subcarrier frequencies.

To estimate the MIMO radar channel response in $ n^\text{th} $ subcarrier, we employ least-squares (LS) estimator \cite{book_kay_estimation_1993} with the precoded symbols that are known at the joint radar transceiver as
\begin{equation}
	\label{eqn_radar_chan_est}
	\radarChanEst_{n} = \boldsymbol{\mathcal{Y}}^{\mathrm{rad}}_{n} \textbf{X}_{n}^{\dagger} \left(\textbf{X}_{n} \textbf{X}_{n}^{\dagger} \right)^{-1}
\end{equation}
where $ \textbf{X}_{n} = \textbf{F}_{n}\textbf{S}_{n}$ is the precoded symbols. However, LS estimator requires the transmitted symbols to be orthogonal (i.e., $ \textbf{X}_{n} \textbf{X}_{n}^{\dagger} = \sqrt{\Ptx} \textbf{I}_{\Ntx}$) to keep the noise component white without amplification, which is achieved with the preamble.  

\textit{Remark}: Although orthogonal transmission achieves the optimal estimation accuracy with LS estimator with the highest degrees of freedom, it disseminates the transmit power omnidirectionally. However, precoding generates a non-orthogonal transmission that prevents the use of LS estimator due to noise amplification. Hence, while transmitting precoded streams, we estimate the radar channel with $\radarChanEst_{n} = \boldsymbol{\mathcal{Y}}^{\mathrm{rad}}_{n} \textbf{X}_{n}^{\dagger} \left(\textbf{X}_{n} \textbf{X}_{n}^{\dagger} \right)^{-1/2}  $ to keep the noise covariance the same. Nevertheless, the radar imaging methods and performance metrics proposed in this section provide insights for the design of optimal precoder.    

As described in Section~\ref{section_system_model}, the composite beampattern of the transmit and receive ULAs is the equivalent of a virtual ULA with $ \Nrx\Ntx $ elements. Thus, we observe the increased virtual apperture with the MIMO radar channel in vector form:
\begin{equation}
	\label{eqn_radar_chan_est_vec}
	\begin{aligned}
		\vectorize\big(\radarChan_{n,p}\big) = \radarChanVec_{n,p} &= \alpha_p \textbf{a}_{\mathrm{rx},n}(R_p,\varTheta_p) \otimes \textbf{a}_{\mathrm{tx},n}(R_p,\varTheta_p) \\
		& = \alpha_p \exponent\big(\!-\!j2\pi f_n 2R_p/c\big) \textbf{u}_n(\varTheta_p),
	\end{aligned}
\end{equation}
where $ \textbf{u}_n(\varTheta) \in \mathbb{C}^{\Nvirt} $ denotes the steering vector of an ULA with $ \Nvirt = \Nrx\Ntx $ antennas whose elements are defined as
\begin{equation}
	\label{eqn_virtual_steer}
	\begin{gathered}
		\textbf{u}_n(\theta)[m] = \exponent\Big(j2\pi f_n m d \sin(\theta)/c  \Big), \\
	\end{gathered}
\end{equation}
where $ d =\lambda/2 $ is the spacing. Hence, with \eqref{eqn_radar_chan_est_vec}, we obtain the response of the virtual ULA for the $ n^{\mathrm{th}} $ subcarrier that provides high resolution for angle estimation. 

Using the vectors defined in \eqref{eqn_radar_chan_est_vec}, we form an observation matrix $ \boldsymbol{\mathcal{H}}  = \big[ \radarChanVec_{-\Nsc/2}, \mydots, \radarChanVec_{\Nsc/2\!\shortminus\!1} \big]^T \in \mathbb{C}^{\Nsc \times \Nvirt} $
whose elements are defined as
\begin{subequations}
	\label{eqn_radar_measurement2}
	\begin{align}
		\boldsymbol{\mathcal{H}}[n ,m]& = \alpha_p\  \exponent\big(\!-\!j2\pi (\fc-\Nsc \fd/2) 2R_p/c\big) \label{eqn_phase1}  \\
	 \times &\ \exponent\big(\!-\!j2\pi n \fd 2R_p/c\big) \label{eqn_phase2}\\
	 \times &\ \exponent\big(j2\pi \fc m {d} \sin(\varTheta_p)/c \big) \label{eqn_phase3}\\
	 \times &\ \exponent\big(j2\pi (n-\Nsc/2) \fd m d \sin(\varTheta_p)/c \big), \label{eqn_phase4} 
	\end{align}
\end{subequations}
The phase distortions are observed due to the reflection from the target as 4 separate terms along frequency axis $ n $ and spatial axis $ m $: (i) a constant phase shift in \eqref{eqn_phase1}, (ii) a linear phase shift along $ n $-axis due to target's range in \eqref{eqn_phase2}, (iii) a linear phase shift along $ m $-axis due to target's angle in \eqref{eqn_phase3}, and (iv) a \textit{coupled} phase term along both axes in \eqref{eqn_phase4}. While \eqref{eqn_phase2} and \eqref{eqn_phase3} are uncoupled across $ n $ and $ m $ axes, \textit{the frequency-space coupling term} in $ \eqref{eqn_phase4} $ reduces the coherency among samples and causes unwanted distortions. 

\subsection{Resolving the Coupling Effect}\label{subsection_coupling}

As highlighted in \eqref{eqn_radar_measurement2}, a frequency-space coupling term \eqref{eqn_phase4} emerges based on target's angle $ \varTheta_p $ and it causes distortions due to a non-linear phase component. In fact, the coupling term is the result of frequency-dependent steering vectors as defined in \eqref{eqn_radar_chan_est_vec}. As the steering vector changes across different subcarriers due to different carrier frequencies, this creates unaligned spatial responses.

The desired response is achieved when steering vectors on different subcarriers have the same carrier frequency $ \fc $, which is not the case due to the OFDM modulation. Therefore, we introduce a scaling factor along $ m $-axis as $ m = \left(\frac{\fc}{\fc + n \fd}\right)m' $ to remove the coupling effect. To apply scaling, we first interpolate samples along spatial-axis $ m $ expressed as
\begin{equation}
	\label{eqn_interp}
	\begin{gathered}
		\begin{aligned}
			\tilde{\mathcal{H}}_n(\tilde{m}) & = \mathrm{interp}\big(\boldsymbol{\mathcal{H}}[n ,:] \big) \\
			& = \varPhi_n \exponent\Big(j2\pi \big(\fc+ n' \fd \big) \tilde{m} {d} \sin(\varTheta_p)/c \Big),
		\end{aligned}
	\end{gathered}
\end{equation}
where $n = 0,\mydots,\Nsc\!\shortminus\!1$, $ n' = n-\Nsc/2$, $ \varPhi_n$ is the constant terms, and $ \mathrm{interp}(.) $ is an interpolation function which can be a sinc or spline interpolator. Now, we can apply the scaling by resampling the interpolated data $ \tilde{\mathcal{H}}_n(\tilde{m}) $ at $ \tilde{m} = \left(\frac{\fc}{\fc + n' \fd}\right)m'$ as
\begin{equation}
	\label{eqn_resampled}
	\begin{gathered}
		\begin{aligned}
			\bar{\mathcal{H}}[n, m']& = \tilde{\mathcal{H}}_n\big(m' \fc/(\fc + n' \fd)\big) \\
			& = \varPhi_n \exponent\Big(j2\pi m' d \sin(\varTheta_p)/c \Big),
		\end{aligned}
	\end{gathered}
\end{equation}
for $n = 0,\mydots,\Nsc\!\shortminus\!1$, and $ m' = 0,\mydots,\Nrx\Ntx\!\shortminus\!1$. In \eqref{eqn_resampled}, the coupling term is eliminated by rescaling the samples along $ m $-axis through interpolation and resampling. Since the rescaling process does not require any prior knowledge regarding target's parameters, it readily mitigates the effects of the coupling term \textit{for multiple targets}.

\subsection{Generating the Radar Image with DFT}\label{subsection_dft_algorithm}

By using the method introduced in Section~\ref{subsection_coupling}, we eliminate the coupling term in \eqref{eqn_phase4}. Henceforth, we show how the range-angle image is generated and target parameters are estimated efficiently without the coupling term by resorting to the discrete Fourier transforms. As shown in \eqref{eqn_radar_measurement2}, $ R_p $ and $ \varTheta_p $ introduce linear phase shifts along $ n $ and $ m $ axes, respectively. Thus, the parameter estimation problem becomes a spectrum analysis problem. 

To determine the rate of change of phase due to $ R_p $, the IDFTs are performed along $ n $-axis for fixed $ m $ values:  
\begin{equation}
\label{eqn_idft_range}
\begin{gathered}
	\begin{aligned}
	\boldsymbol{\Omega}'[r,m] & = \sum_{n=0}^{\Nsc\!\shortminus1}  \bar{\boldsymbol{\mathcal{H}}}[n,m] \exponent\big(j2\pi r n /\Nsc\big), \\ 
	&= \varPhi_m \sum_{n=0}^{\Nsc\!\shortminus1} \exponent\big(\!-\!j2\pi n \fd 2R_p/c\big) \exponent\big(j2\pi r n /\Nsc\big), 
	\end{aligned}
\end{gathered}
\end{equation}
where $ \varPhi_m $ is the constant values in $ \bar{\textbf{S}}[:,m] $. With the IDFT, $ \boldsymbol{\Omega}'[:,m] $ contains the range profiles observed by each transmit-receive antenna pair with a peak where the exponent terms cancel each other. With the index of the peak denoted by $ \hat{r} $, the range estimate $ \hat{R}_p $ is derived as
\begin{equation}
	\label{eqn_range_est}
	\hat{R}_p = \frac{\hat{r} c}{2\Nsc \fd},\ \ \hat{r} = 0,\mydots,\Nsc\!\shortminus\!1. 
\end{equation}
Correspondingly, the DFTs are performed on $ \boldsymbol{\Omega}'[r,m] $ along the $ m $-axis for fixed $ n $ values to estimate the rate of change of phase due to $ \varTheta_p $:
\begin{equation}
	\label{eqn_dft_angle}
	\begin{gathered}
		\begin{aligned}
			\boldsymbol{\Omega}&[r,\theta]  = \sum_{m=0}^{\Nvirt\!\shortminus\!1}  \boldsymbol{\Omega}'[r,m] \exponent\big(j2\pi m \theta /\Nvirt \big), \\ 
			&= \varPhi_r \sum_{m=0}^{\Nvirt\!\shortminus\!1} \exponent\big(j2\pi \fc m {d} \sin(\varTheta_p)/c \big) \exponent\big(j2\pi m \theta /\Nvirt \big), 
		\end{aligned}
	\end{gathered}
\end{equation}
where $ \varPhi_r $ is the constant values in $ \boldsymbol{\Omega}'[r,:] $. After performing the DFTs, $ \boldsymbol{\Omega} $ contains a single peak, where the exponents in \eqref{eqn_dft_angle} cancel each other. With the index of the peak at $ \hat{\theta} $ along the $ \theta $-axis, the angle estimate $ \hat{\varTheta}_p $ is given as
\begin{equation}
	\label{eqn_angle_est}
	\hat{\varTheta}_p = \sin^{-1}\left(\frac{2 \hat{\theta}}{\Nvirt}\right),\ \ \hat{\theta} = -\Nvirt/2,\mydots,\Nvirt/2\!\shortminus\!1,
\end{equation}
assuming the zero-frequency is placed at the center of the spectrum. While the DFTs offer low computation complexity for the spectral estimation with the fast Fourier transform (FFT) algorithm, they generate a quantized spectrum with a wide mainlobe and sidelobes as described in \cite[Chapter~14]{book_richards_radar_principles_2010} which can impair \textit{multiple target separation} capabilities. 

After performing the DFTs along both axes, the resulting data matrix $ \boldsymbol{\Omega} \in \mathbb{C}^{\Nsc \times \Nvirt} $ is the complex range-angle image that contains a peak at $ (\hat{r}, \hat{\theta})$ on the quantized grid. Therefore, the target detection and the parameters estimation using DFT-based processing suffer from quantization error and masking due to the nulls and sidelobes. To mitigate the quantization error, a finer grid for the range-angle image can be achieved by interpolation of the quantized spectrum as studied in \cite{book_richards_radar_principles_2010}. A sinc (i.e., bandlimited) interpolation can be performed by zero-padding the data at the cost of increased computational complexity due to the longer DFTs. Furthermore, applying window function on the measurement data attenuates the sidelobes due to spectral leakage at the expense of lower resolution due to widened mainlobe. The proposed processing method is outlined in Algorithm~\ref{algo_dft}.

\begin{algorithm}[!t]
	\DontPrintSemicolon
	\SetKwInOut{Input}{input}
	\SetKwInOut{Output}{output}
	\KwIn{$\boldsymbol{\mathcal{H}} \in \mathbb{C}^{\Nsc\times\Nvirt}$: the observation matrix, \\
		$ \Dsc $ and $ \Dvirt $: the length of DFT operations\\
	}
	\KwOut{$ |\boldsymbol{\Omega}| \in \mathbb{R}^{\Dsc\times\Dvirt}$: the range-angle image}
	$ \bar{\boldsymbol{\mathcal{H}}} \gets $ Eliminate the coupling in $ \boldsymbol{\mathcal{H}} $ using \eqref{eqn_interp} and \eqref{eqn_resampled}\\
	
	$ \bar{\boldsymbol{\mathcal{H}}}_{\mathrm{win}} \gets $ Apply window function on $ \bar{\boldsymbol{\mathcal{H}}} $ along both axis \\
	
	$ \bar{\boldsymbol{\mathcal{H}}}_{\mathrm{pad}} \gets $ Zero-pad $ \bar{\boldsymbol{\mathcal{H}}}_{\mathrm{win}} $ to match the size of $ \Dsc \times \Dvirt $\\

	$ \boldsymbol{\Omega}'\gets $ Perform IDFTs along first axis of $ \bar{\boldsymbol{\mathcal{H}}}_{\mathrm{pad}} $ as in \eqref{eqn_idft_range}\\
	
	$ \boldsymbol{\Omega} \gets $ Perform DFTs along second axis of $ \boldsymbol{\Omega}' $ as in \eqref{eqn_dft_angle}\\
		
	\Return{$ |\boldsymbol{\Omega}| $}\;
	\caption{MIMO-OFDM Radar Imaging Method}
	\label{algo_dft}
\end{algorithm}

\subsection{Performance Metrics}\label{subsection_parameters}
The target's range and angle estimates are obtained with the DFTs in \eqref{eqn_idft_range} and \eqref{eqn_dft_angle} which generate the complex-valued range-angle image as a 2D spectrum analysis. As shown in \cite{complex_crlb_1998}, DFT-based frequency estimation with interpolation achieves an estimation performance that is close to the maximum likelihood (ML) estimator whose variance is lower bounded by the Cramer–Rao lower bound (CRLB). Assuming the coupling term in \eqref{eqn_radar_measurement2} is removed and target's are resolved in the radar image, we can adapt the CRLB on the variance of frequency estimation for the range and angle parameters with the observation matrix in \eqref{eqn_resampled}. Based on CRLB derived for the variance of frequency estimation with a complex sinusoid in \cite{complex_crlb_1998}, we derive the CRLBs for the variances of range and angle estimates of the $p^{\text{th}} $ target as
\begin{equation}
		\setlength{\jot}{5pt}
   	\label{eqn_crlb_range}
   		\begin{gathered}
   	\underline{\sigma}_{R,p}^2 =  \frac{3 c^2}{8 \pi^2 \fd^2 (\Nsc^2 - 1) \snrradar_p 	 \mathrm{P}(\varTheta_p)},  \\
	\underline{\sigma}_{\varTheta,p}^2 =  \frac{6 c^2}{\pi^2 \fc^2 \lambda^2 \cos^2(\varTheta_p) (\Nvirt^2 - 1) \snrradar_p 	 \mathrm{P}(\varTheta_p)} , 
  		\end{gathered}
\end{equation}
where $  \mathrm{P}(\varTheta_p) $ is the average transmit power directed towards the target which is defined as 
\begin{equation}
	\label{eqn_tx_power}
	\mathrm{P}(\varTheta_p) = \sum_{n=0}^{\Nsc\!\shortminus1} \left|\textbf{u}_{\mathrm{tx},n}^{\dagger}(\varTheta_p) \textbf{F}_n \textbf{F}_n^{\dagger} \textbf{u}_{\mathrm{tx},n}(\varTheta_p)\right|/\Nsc ,
\end{equation}
and $ \textbf{u}_{\mathrm{tx},n} $ is the steering vector of the transmit ULA.
The signal-to-noise ratio (SNR) of the target is defined as
\begin{equation}
	\label{eqn_snr_radar}
	\snrradar_p = \frac{\Nvirt\Nsc\Nsym|\alpha_p|^2 }{\sigma^2_{\mathrm{rad}}},
\end{equation}

\subsubsection*{\textbf{Positioning Accuracy}}
While we derive separate CRLBs for range and angle estimation in \eqref{eqn_crlb_range}, we wish to obtain single positioning accuracy metric for target $ p $ in Cartesian coordinates. As proposed in \cite[Chapter~3.8]{book_kay_estimation_1993}, we change variables as $ x_p = R_p \sin (\varTheta_p) $ and $ y_p = R_p \cos (\varTheta_p) $ to obtain the CRLBs for the variances of Cartesian coordinate estimates as 
\begin{equation}
	\setlength{\jot}{5pt}
	\label{eqn_crlb_xy}
	\begin{gathered}
		\begin{aligned}
			\underline{\sigma}_{x,p}^2 & = (\partial x_p/\partial R_p)^2 \underline{\sigma}_{R,p}^2 + (\partial x_p/\partial \varTheta_p)^2 \underline{\sigma}_{\varTheta,p}^2, \\ 
			\underline{\sigma}_{y,p}^2 & = (\partial y_p/\partial R_p)^2 \underline{\sigma}_{R,p}^2 + (\partial y_p/\partial \varTheta_p)^2 \underline{\sigma}_{\varTheta,p}^2.
		\end{aligned}
	\end{gathered}
\end{equation}
Based on \eqref{eqn_crlb_xy}, the positioning accuracy of the target is defined as 
\begin{equation}
	\label{eqn_crlb_pos}
	\delta^2_{\mathrm{pos},p} = \frac{1}{\underline{\sigma}_{x,p}^2+\underline{\sigma}_{y,p}^2} = \frac{1}{\underline{\sigma}_{R,p}^2+\underline{\sigma}_{\varTheta,p}^2 R_p^2},
\end{equation}
which is the inverse of positioning CRLB as a function of target's parameters $ \{\alpha_p, R_p, \varTheta_p\}  $ and precoders $ \textbf{F}_n $.

With the DFTs, the resulting spectrum is the superposition of sinc functions based on the spectral components. Hence, the step sizes in \eqref{eqn_range_est} and \eqref{eqn_angle_est} determines the imaging resolution of the sinc function \cite{book_richards_radar_principles_2010}. Particularly, the range resolution denoted by $ \rangeRes $ and angle resolution denoted by $ \angleRes $ are derived as
\begin{equation}
	\label{eqn_resolutions}
	\rangeRes = \frac{c}{2 B}\: \text{ and }\:  \angleRes = \frac{2}{\Nvirt},
\end{equation}
respectively. While the range resolution is determined solely by the total bandwidth $ B = \Nsc \fd$ of the signal, the angle resolution is determined solely by the number of antennas in virtual ULA $ \Nvirt = \Nrx \Ntx $. Moreover, the maximum unambiguous range is determined by the maximum index in \eqref{eqn_range_est} as $\rangeMax = (\Nsc c)/(2 B)$, which is the range limit due to the ambiguity of the spectrum estimated with limited samples attained from subcarriers. 

In addition to the maximum unambiguous range $ \rangeMax $, the cyclic prefix duration also determines the ISI-free maximum range which is defined as 
\begin{equation}
	\label{eqn_maximums}
	\rangeMax^{(ISI)} = \frac{\Tgi c}{2}
\end{equation}
Therefore, to prevent ambiguity and ISI in reflected OFDM symbols the maximum range requirement of the radar should be lower than the minimum of $ \rangeMax $ and $ \rangeMax^{(ISI)} $. 

\section{Joint Precoder Design} \label{section_precoder}

The joint transceiver periodically transmits NDPs when no communication link is established for both MIMO radar imaging and channel sounding with an orthogonal transmission as described in Section~\ref{section_system_model} and \ref{section_range_angle}. Once the CSI that contains $\{\hat{\textbf{h}}^{\mathrm{com}}_{n,q}, \hat{\sigma}^2_{\mathrm{com},q}\}_{q\in\Psi_{\mathrm{com}}} $ is received from $ \Mrx $ communication receivers, the joint transceiver can transmit independent streams by precoding the data. However, designed precoder can degrade the target tracking and imaging performance of radar during the data transmission. 

With the prior NDP transmissions, the joint transceiver also acquires the parameters of $ \Mtarget $ targets denoted by $\{\hat{\alpha}_p, \hat{R}_p,\hat{\varTheta}_p\}_{p \in\Psi_{\mathrm{rad}}} $ that are resolved in the radar image. Instead of resorting to conventional precoding methods, we propose a joint precoder design problem to transmit data by meeting \textit{the minimum SINR requirement} while illuminating previously detected target with \textit{high positioning accuracy} defined in \eqref{eqn_crlb_pos} which is the inverse of \textit{positioning CRLB}.

For this problem, our objective is to allocate the maximum transmit powers on the targets of interest as defined in \eqref{eqn_tx_power} to achieve the same level of positioning accuracies for all targets that is achieved by a $\max$-$\min$ optimization. Based on the performance metrics for radar in \eqref{eqn_crlb_pos} and communication in \eqref{eqn_comm_sinr}, we formulate joint precoder design problem for $ n^{\mathrm{th}} $ subcarrier as
\begin{equation}
	\begin{aligned}
		& \underset{ \{\textbf{f}_{i,n}\}_{i=1}^{\Ntx} }{\text{max}} && \underset{ p \in\Psi_{\mathrm{rad}} }{\text{min}}\ 
		  \omega_p\left|\textbf{u}_{\mathrm{tx},n}^{\dagger}(\hat{\varTheta}_p) \textbf{R}_{F,n} \textbf{u}_{\mathrm{tx},n}(\hat{\varTheta}_p)\right|, \\
		&\ \  \text{s.t.}
		& \conn_1&:\textbf{R}_{F,n} = \sum_{i=1}^{\Ntx} \textbf{f}_{i,n}\textbf{f}_{i,n}^{\dagger} \succeq 0, \\
		&& \conn_2&: \trace \left(\textbf{R}_{F,n}\right) \leq \Ptx, \\
		&& \conn_3&:  \SINRcomm_{n,q}\left(\textbf{f}_{q,n}\right) \geq \eta_q, \ \forall q\in\Psi_{\mathrm{com}}, \\ 
		&& \conn_4&: \left| \textbf{Q}_{F,n}[k,l]\right| \leq  \gamma_{\mathrm{cor}}, \ \forall l\neq k, %\textbf{Q}_{F,n}[k,k]
	\end{aligned} \tag{P0} \label{opt_tx_design}
\end{equation}
where $ \omega_p $ is the weighting parameter based on the constants in $\delta^2_{\mathrm{pos},p} $ for fair power allocation, $ \textbf{Q}_{F,n} = \boldsymbol{\mathcal{U}}_{\mathrm{ang}}^\dagger \textbf{R}_{F,n} \boldsymbol{\mathcal{U}}_{\mathrm{ang}} $ is the correlation matrix in angular domain. Also, $ \boldsymbol{\mathcal{U}}_{\mathrm{ang}}  \in \mathbb{C}^{\Ntx \times \Nang}$ contains the concatenated steering vectors of $ \Nang $ angles of interest based on the directions of targets and receivers. As studied in \cite{li2009mimo}, $ \conn_4 $ is used to prevent high angular correlation between different directions when generating transmit beams as required for accurate MIMO radar processing. Since a communication receiver can be detected as a target at the same angle, $ \Nang \leq \Mrx + \Mtarget $.

The problem in \eqref{opt_tx_design} generates directional transmit beams for the targets to maximize the minimum positioning accuracy with a transmit power limit in $ \conn_2 $, minimum SINR requirements for communication receivers in $ \conn_3 $, and angular correlation constraint in $ \conn_4 $. To recast the problem in \eqref{opt_tx_design} as a convex semidefinite programming (SDP) problem, we first change the variable as $ \textbf{R}_{i,n} = \textbf{f}_{i,n}\textbf{f}_{i,n}^{\dagger} $ and reformulate the SINR constraint $ \conn_3 $ as
\begin{equation}
\label{eqn_sinr_sdp}
\begin{gathered}
\begin{aligned}
\SINRcomms_{n,i}\left(\textbf{R}_{i,n}\right) =  (1 + 1&/\eta_i)\trace\left( \textbf{R}_{i,n} \hat{\textbf{H}}^{\mathrm{com}}_{n,i} \right) \\
& - \trace\left( \textbf{R}_{F,n} \hat{\textbf{H}}^{\mathrm{com}}_{n,i} \right) - \hat{\sigma}^2_{\mathrm{com},i},
\end{aligned}
\end{gathered}
\end{equation}
where $ \hat{\textbf{H}}^{\mathrm{com}}_{n,q} = \hat{\textbf{h}}^{\mathrm{com}}_{n,q} {\hat{\textbf{h}}^{\mathrm{com} \dagger}}_{n,q} $. By omitting the non-convex rank-one constraints $ \rank(\textbf{R}_{i,n}) = 1 $, we obtain the convex SDP version of the original problem \eqref{opt_tx_design} as 
\begin{equation}
	\begin{aligned}
		& \underset{\{\textbf{R}_{i,n}\}_{i=1}^{\Ntx}, \tau}{\text{max}} && \tau, \\
		& \ \ \ \  \text{s.t.}
		& \conn_2&, \conn_4, \\
		&& \conn_1&:\textbf{R}_{F,n} = \sum_{i=1}^{\Ntx} \textbf{R}_{i,n} \succeq 0, \\
		&& \conn_3&: \SINRcomms_{n,q}\left(\textbf{R}_{q,n}\right) \geq 0,  \ \ \forall q\in \Psi_{\mathrm{com}}, \\ 
		&& \conn_5&: \omega_p \trace\left(\textbf{R}_{F,n} \textbf{U}_{\mathrm{tx},n}(\hat{\varTheta}_p) \right) \geq \tau, \ \ \forall p\in\Psi_{\mathrm{rad}},
	\end{aligned} \tag{P-SDP} \label{opt_tx_design_sdp}
\end{equation}
where $ \tau $ is an auxiliary variable to replace the $ \min $ optimization, $ \textbf{U}_{\mathrm{tx},n}(\hat{\varTheta}_p) = \textbf{u}_{\mathrm{tx},n}(\hat{\varTheta}_p)\textbf{u}^{\dagger}_{\mathrm{tx},n}(\hat{\varTheta}_p) $. Also, $ \conn_2 $, and $ \conn_4 $ in  \eqref{opt_tx_design_sdp} are the same constraints from \eqref{opt_tx_design}. 

 Since \eqref{opt_tx_design_sdp} is a convex SDP problem, we can solve it in polynomial time to obtain $ \{\textbf{R}_{i,n}^{\star}\}_{i=1}^{\Ntx} $ with interior-point algorithms that are available in convex optimization toolbox CVX \cite{cvx}. The precoder vectors of the original problem \eqref{opt_tx_design} are obtained as $ \textbf{f}_{i,n}^\star = \sqrt{\lambda_{i,n}}\textbf{v}_{i,n}$ where $ \lambda_{i,n} $ and $ \textbf{v}_{i,n} $ denote the largest eigenvalue and corresponding eigenvector of $ \textbf{R}_{i,n}^{\star} $, respectively. When $ \textbf{R}_{i,n}^{\star} $ is a rank-one matrix, $ \textbf{f}_{i,n}^\star $ is the optimal solution. Nevertheless, $ \textbf{f}_{i,n}^\star $ is still the best rank-one approximation of $ \textbf{R}_{i,n}^{\star} $, if it is not rank-one. 
 \begin{figure}[!b]
	\centering
	\includegraphics[width=\columnwidth]{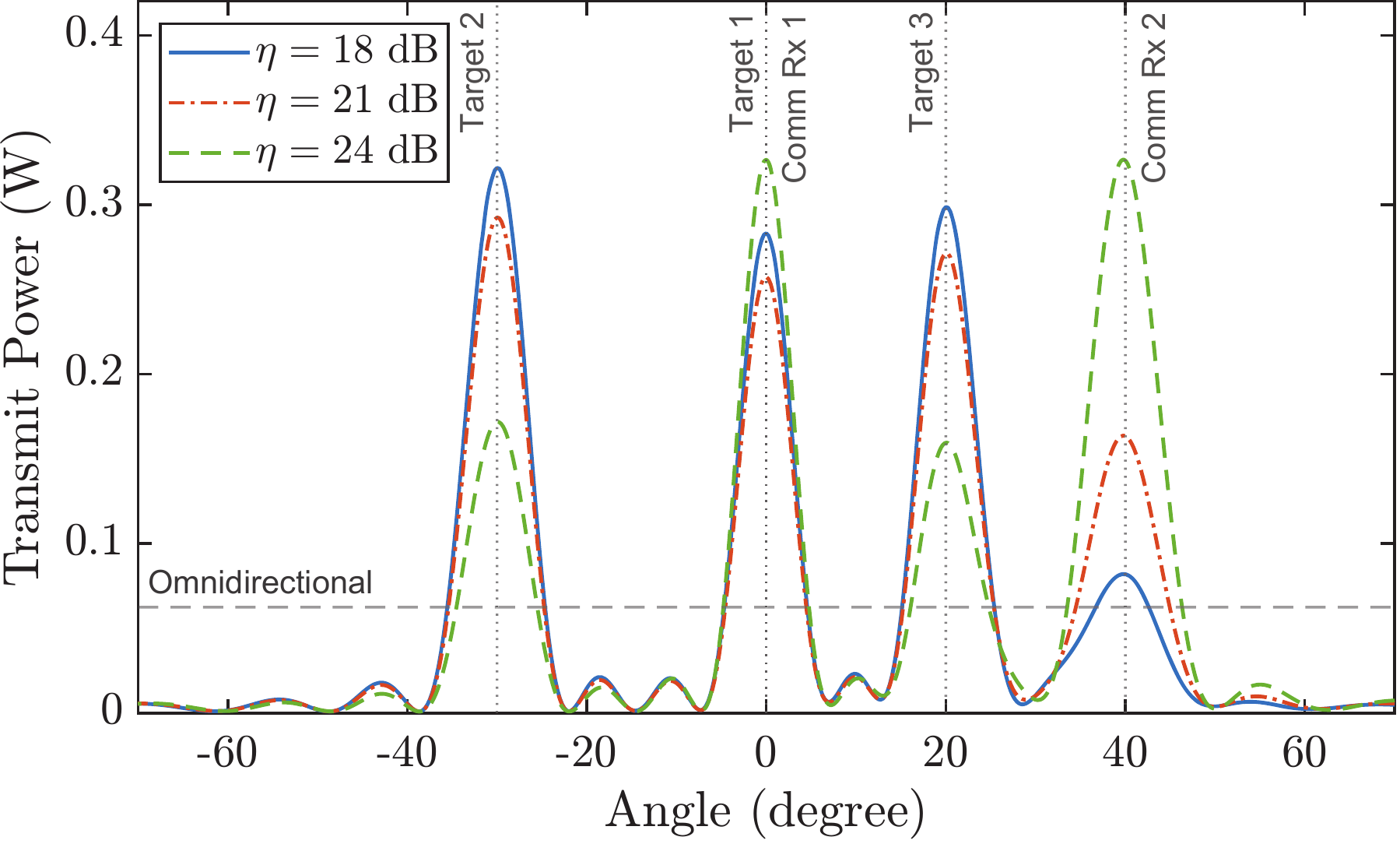} 
	\caption{The transmit pattern of the designed precoder for different communication SINR constraints.}
	\label{fig_pattern_opt}
\end{figure}

\begin{table}[!t]
	\caption{System parameters for the radar simulation.}
	\centering
	\resizebox{\columnwidth}{!}{
		\begin{tabularx}{\columnwidth}{c C c}
			\hline \hline
			\textbf{Symbol} & \textbf{Parameter} & \textbf{Value} \\ \hline\hline
			$f_c$ & Carrier frequency & 76.5 GHz \Tstrut\\
			$B$ & Bandwidth & 160 MHz \\
			$ \Nsc $ & Number of subcarriers & 64 \\ 
			$ \Ntx $ & Number of transmit antennas & 16 \\ 
			$ \Nrx $ & Number of receive antennas & 4 \\ 
			\hline
			$\Tdft$ & OFDM symbol duration  & 0.40 $ \upmu $s \Tstrut\\
			$\Tgi$  & Cyclic prefix duration & 0.40 $ \upmu $s\\
			$\Tsym$ & Total OFDM symbol duration & 0.80 $ \upmu $s\\ \hline
			%			$ \rangeRes $ & Range resolution & 0.94 m \Tstrut\\
			%			$ \angleRes $ & Angular resolution & 1.80$^{\circ}$ \\
			%			$ \rangeMax $ & Unambiguous maximum range & 60 m \\
			$\Ptx $ & Total transmit power  & 30 dBm \Tstrut \\
			$F_{\mathrm{n}}$ & Noise figure & 15 dB \\ 
			\hline
		\end{tabularx}
	}
	\label{table_parameters}
\end{table}

\section{Simulation and Numerical Results}

In this section, we evaluate the performance of the proposed range-angle imaging and joint precoder design methods through simulations in MATLAB. In the simulations, we consider a vehicular scenario in which a joint transceiver vehicle receives CSI from two communication receivers with single antennas as responses to the previous NDP transmission. The MIMO-OFDM signal is generated with a fixed preamble and quadrature phase shift keying (QPSK) modulated data and radar streams. The preamble is generated with a Hadamard mapping matrix that consists of orthogonal sequences to generate an omnidirectional transmit pattern which radiates the transmit power equally as $ \Ptx/\Ntx $. The system parameters of the joint transceiver are summarized in Table~\ref{table_parameters}.

The communication receivers are located 25 m away at azimuth angles of 0$^{\circ}$ and 40$^{\circ}$. The communication channels contains line-of-sight paths whose gains are computed with receive gain of 20 dB and free-space path loss. The illuminated region also contains of three vehicles with the reflectivity of 20 dBsm that are located 25 m at azimuth angles of 0$^{\circ}$, -30$^{\circ}$, and 20$^{\circ}$, which correspond to (25,0), (21.7, -12.5), and (23.5, 8.6) in Cartesian coordinates, respectively. Hence, the joint transceiver also tracks the targets that are modeled as point-scatterers by estimating their range, angle and gain values with the NDP transmissions. The propagation paths due to reflections from targets are computed for all transmit-receive antenna pairs as defined in \eqref{eqn_rx_signal} to obtain the reflected passband signal. Also, we note that every communication receiver may not be detected by radar (e.g., the receiver at 40$^{\circ}$ angle is not detected) since reflected signals experience high attenuation.

\begin{figure}[!t]
	\centering
	\includegraphics[width=\columnwidth]{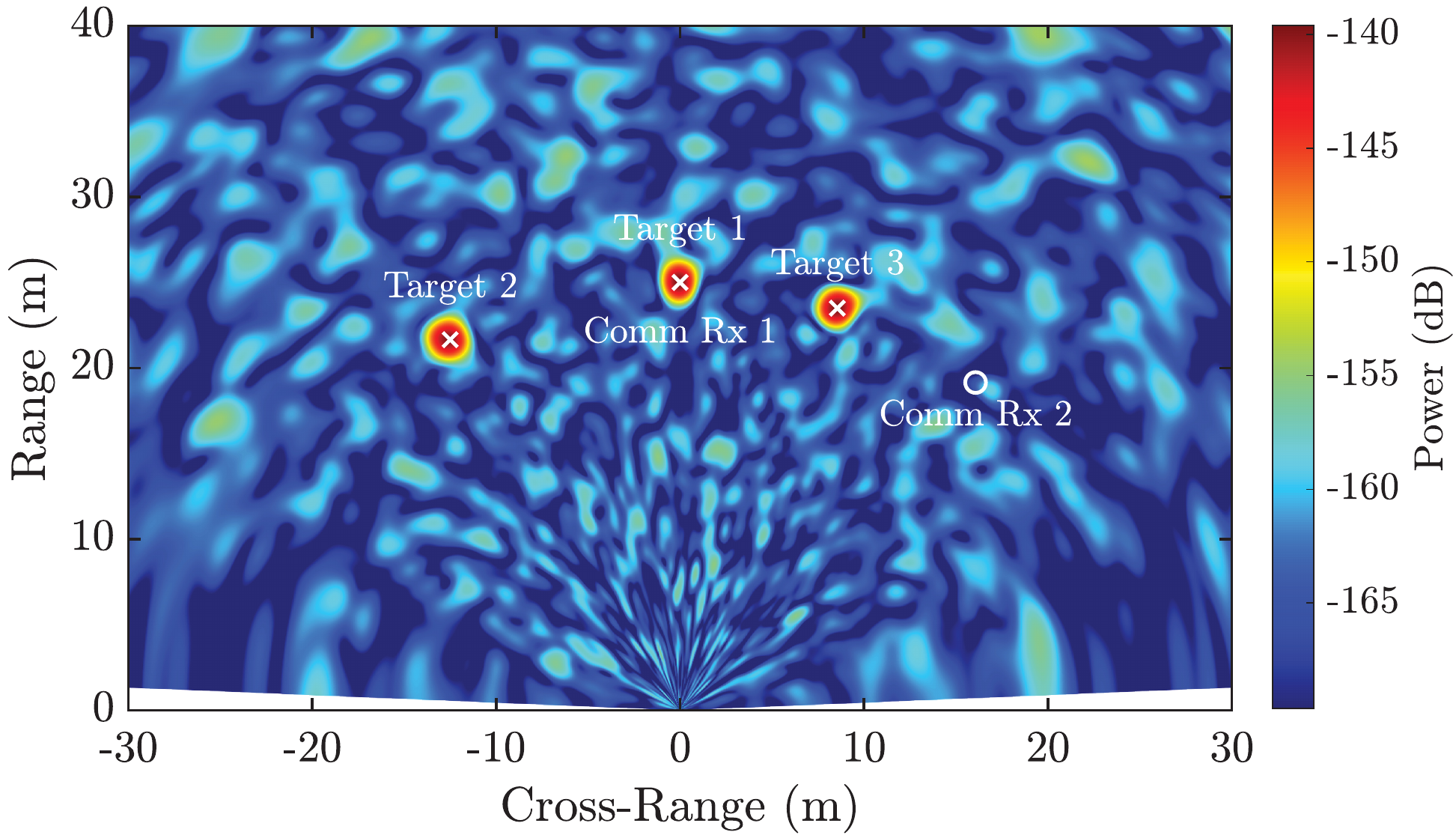} 
	\caption{The radar image generated with NDP where markers denote the actual position of the targets.}
	\label{fig_dft}
\end{figure}

\begin{figure}[!b]
	\centering
	\includegraphics[width=\columnwidth]{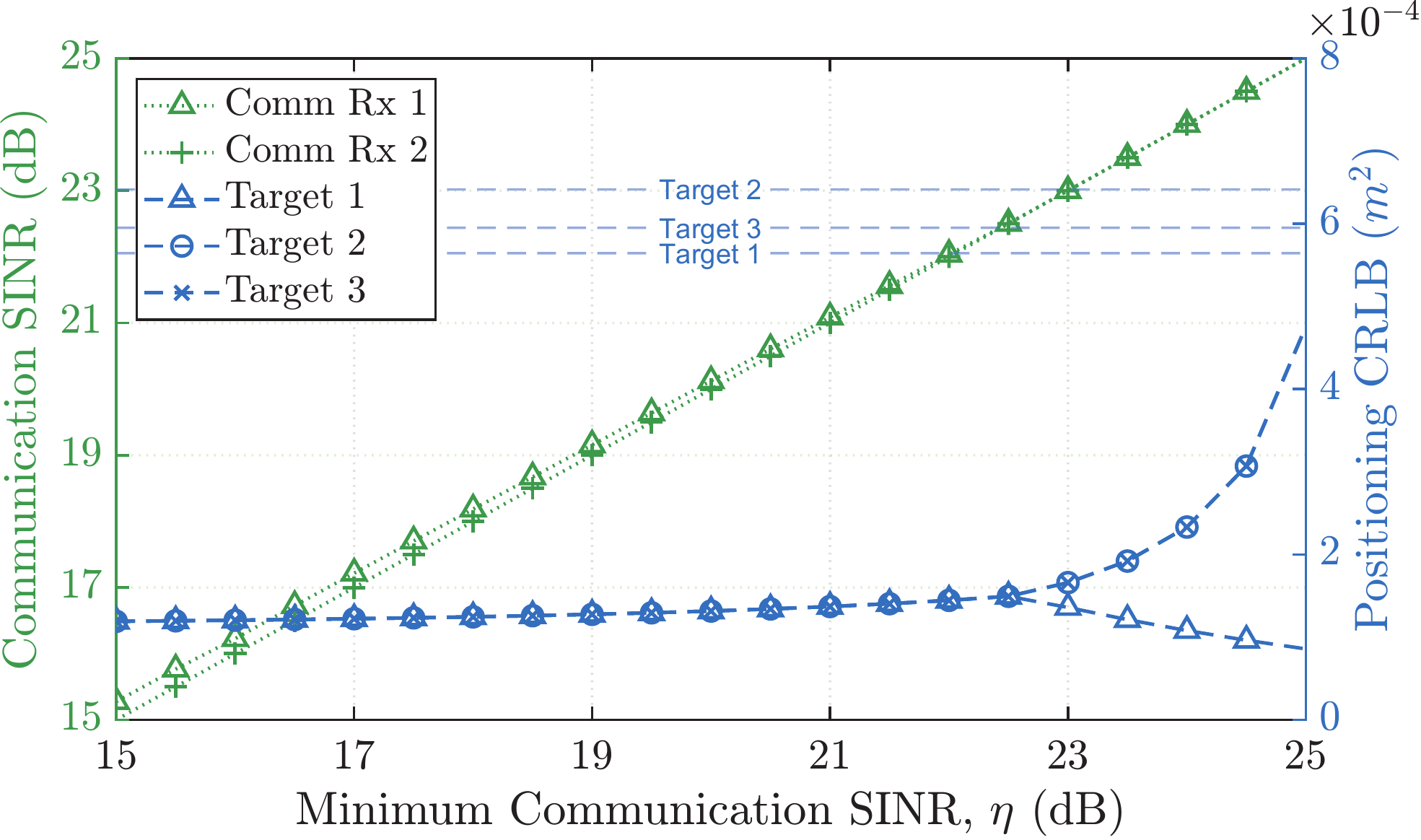} 
	\caption{Achieved communication SINRs and positioning CRLBs. (Vertical lines: the CRLBs with omnidirectional transmission.)}
	\label{fig_results}
\end{figure}

By leveraging the estimates of targets' parameters and the received CSI, the joint transceiver designs the optimal precoder with \eqref{opt_tx_design_sdp} where the same minimum SINR constraints are used for the communication receivers. For radar imaging with the precoded symbols, $ \Nsym = \Ntx $ symbols are used to generate the range-angle image. Moreover, we assume that the targets are quasi-stationary considering the total transmission duration is only $\Ntx\Tsym = 12.8\ \upmu $s. The joint transceiver and communication receivers have noise figure of $ F_{\mathrm{n}} =$ 15 dB that corresponds to total noise power of $ -106.93 $ dB.

With the designed precoder, the transceiver performs simultaneous spatial data multiplexing and target tracking. In Fig. \ref{fig_pattern_opt} and \ref{fig_results}, the achieved transmit patterns and performances are shown for different SINR constraints $ \eta $. With the lower SINR requirements (e.g., $ \eta < 22 $dB) for communication, the most of the transmit power is directed towards the targets while attaining the same positioning accuracy as shown in Fig.~\ref{fig_results}. Hence, the largest power is allocated to $ \mathsf{Target\ 2} $ due to its wider angle where the positioning accuracy is the worst as given in \eqref{eqn_crlb_pos}. As the SINR requirement increases, the more power is directed towards the communication receivers. Since $ \mathsf{Comm\ Rx\ 1} $ and $ \mathsf{Target\ 1} $ are located at the same angle, high SINR requirement leads to higher transmit power steered towards $ \mathsf{Target\ 1} $ compared to other targets as observed in Fig. \ref{fig_results} for $ \eta > 22 $ dB. However, sustaining two high data rate communication links lowers the tracking accuracy substantially for \textit{non-receiver targets}. Nevertheless, the beamforming gains and the re-use of a beam allow better positioning accuracies compared to omnidirectional transmission.

\begin{figure}[!t]
	\centering
	\includegraphics[width=\columnwidth]{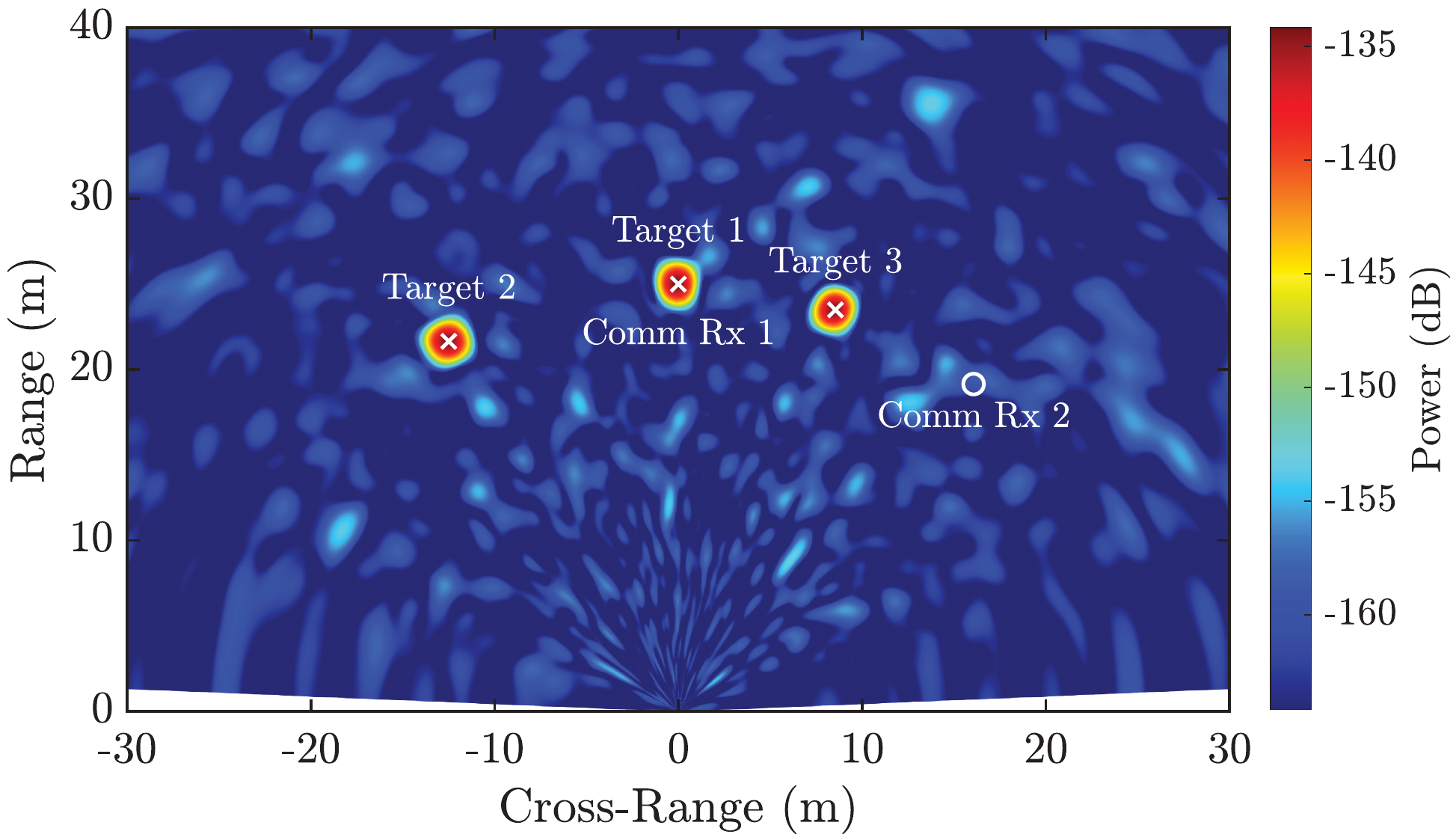} 
	\caption{The radar image generated with the precoded symbols with SINR requirement of $ \eta = 21$ dB.}
	\label{fig_precoder_imag}
\end{figure}

In Fig. \ref{fig_dft} and \ref{fig_precoder_imag}, the radar images that are obtained with Algorithm~\ref{algo_dft} are shown in Cartesian coordinates for NDP and precoded symbols with $ \eta = 21 dB $, respectively. Although the designed joint precoder generates coherent signals that lower the waveform diversity, it generates higher beamforming gain with imaging capability as shown with the higher received power levels. While some power is allocated for spatial multiplexing of data, the proposed joint precoder design methods allows us to reuse common beams and track the targets of interest with high accuracy.

\section{Conclusion}

In this work, we explored the use of MIMO-OFDM waveform for efficient and high resolution radar imaging for the joint automotive radar-communication networks. By exploiting the virtual MIMO radar processing and orthogonal preamble structure, we proposed a DFT-based range-angle imaging method and derived its performance metrics for positioning accuracy. Based on the performance metrics, we proposed a joint precoder design method that enables simultaneous spatial multiplexing for multi-user MIMO communication and high accuracy target tracking.

\bibliographystyle{IEEEtran}
\bibliography{references} 

\end{document}